\documentclass[11pt, conference]{IEEEtran} 
\usepackage{graphics}
\usepackage{amssymb}
\usepackage{amscd}
\usepackage{dsfont}
\usepackage{amsmath}
\usepackage{epsfig}
\usepackage{psfrag}
\usepackage{rotating}
\usepackage{amsmath}
\usepackage{amsfonts}
\usepackage{url}
\usepackage{color}

\newtheorem{theorem}{Theorem}

\newtheorem{definition}[theorem]{Definition}
\newtheorem{proposition}[theorem]{Proposition}

\newcommand{\ind}[1]{\mathds{1}_{\left\lbrace #1 \right\rbrace}}
\newcommand{\expected}[2]{\mathds{E}_{#2}\left[ #1 \right]}

\newcommand{\SNR}{\scriptstyle\mathrm{SNR}}

\newcommand{\bs}{\boldsymbol}
\newcommand{\mc}{\mathcal}

\newcommand{\defn}{\stackrel{\triangle}{=} }
\newcommand{\ds}{\displaystyle}

\newcommand{\diag}{\mathrm{diag}}

\newcommand{\pr}[1]{\mathrm{Pr}\left(#1\right)}

\newcommand{\channelset}{\left\lbrace g_{k,s}\right\rbrace_{\forall (k,s) \in \mathcal{K} \times \mathcal{S}}}

\newcommand{\GameNF}{\mathcal{G} = \left(\mathcal{K}, \left\lbrace\mathcal{A}_k \right\rbrace_{k \in \mathcal{K}},\left\lbrace u_{k}\right\rbrace_{ k \in \mathcal{K}}\right)}
\newcommand{\GameNFtilde}{\mathcal{G}' = \left(\mathcal{K}, \left\lbrace\mathcal{A}_k \right\rbrace_{k \in \mathcal{K}},\left\lbrace \phi\right\rbrace_{ k \in \mathcal{K}}\right)}

\renewcommand{\baselinestretch}{.816}

\begin{document}
\title{On the Fictitious Play and Channel Selection Games}
\author{\IEEEauthorblockN{S.~M. Perlaza}
\IEEEauthorblockA{Orange Labs - Paris,\\
France Telecom R\&D, France.\\
medina@lss.supelec.fr
}
\and
\IEEEauthorblockN{H. Tembine and S. Lasaulce}
\IEEEauthorblockA{Laboratoire des Signaux et Syst\`{e}mes.\\
CNRS - SUPELEC - Univ. Paris Sud, France.\\
$\lbrace$tembine, lasaulce$\rbrace$@lss.supelec.fr
}
\and
\IEEEauthorblockN{V.~Quintero-Florez}
\IEEEauthorblockA{Telecommunications Dept.\\
Universidad del Cauca,  Colombia.\\
vflorez@unicauca.edu.co}
}
\maketitle
\begin{abstract}
Considering the interaction through mutual interference of the different radio devices, the channel selection (CS) problem in decentralized parallel multiple access channels can be modeled by strategic-form games. Here, we show that the CS problem is a potential game (PG) and thus the fictitious play (FP) converges to a Nash equilibrium (NE) either in pure or mixed strategies. Using a $2-$player $2-$channel game, it is shown that convergence in mixed strategies might lead to cycles of action profiles which lead to individual spectral efficiencies (SE) which are worse than the SE at the worst NE in mixed and pure strategies. Finally, exploiting the fact that the CS problem is a PG and an aggregation game, we present a method to implement FP with local information and minimum feedback.
\end{abstract}

\section{Introduction}\label{SecIntroduction}

In recent literature in wireless communications (see \cite{Scutari-2006, Perlaza-gamecomm-09, Perlaza-TSP-09b}), it has been shown that channel selection (CS) problems in decentralized parallel multiple access channels (MAC), where each transmitter is interested in its own spectral efficiency, can be modeled by strategic-form games. More importantly, it has been shown in \cite{Perlaza-gamecomm-09} that CS problems in MAC are finite potential games (PG) \cite{Monderer-Shapley-1996}. In general, finite potential games belong to the class of games for which the existence of at least one Nash equilibrium (NE) is ensured \cite{Monderer-Shapley-1996}. Moreover, it has been shown that the iterative best-response dynamics (BRD) and fictitious play (FP) both converge to Nash equilibrium in potential games \cite{Monderer-Shapley-1996b}. The BRD in CS problems has been studied in \cite{Perlaza-TSP-09b}. In this paper, we exclusively focus on the case of FP in its original version \cite{Brown-1951}. Here, it is shown that when several NE exists in the CS game, which is often the case at high signal-to-noise ratio, FP might converge to (strictly) mixed strategies and cycles of action profiles might be observed. This fact rises the main question to be answered by this paper: is the empirical measure of the frequency of each player's actions a good metric to evaluate convergence? Here, we use a $2-$player $2-$action game to show that the actions or mixed strategies corresponding to an NE are never played even though a convergence of the empirical frequencies of each player is observed. Surprinsingly, cycles of action profiles might lead players to achieve expected utilities which are worse than the worst expected utilities at NE in pure and mixed strategies. 

\noindent
Finally, we also show  that the CS problem is an aggregation game \cite{Rausser-2003}. Here, the utility achieved by a given player does not depend directly on the actions of all the players but on the actions of the given player and a linear combination of all the players' actions. Interestingly, such a linear combination of actions in the CS problem in MAC, is just the multiple access interference (MAI) seen at the receiver. Thus, the MAI observed over each channel at the receiver can be fed back to the transmitters through a common signaling message. Using this message, it is shown that the FP can be implemented by relying on the fact that transmitters are able to obtain an estimate of their own channels and to calculate their utilities based on the above mentioned signaling message.

\noindent
This paper is organized as follows. In Sec. \ref{SecModels}, we formalize the CS problem in parallel MAC and formulate the corresponding strategic-form game. In Sec. \ref{SecExistingResults}, we present recent results on the existence and multiplicity of the NE in the CS game. In Sec. \ref{SecFictitiousPlay}, we introduce the FP originally introduced in \cite{Brown-1951} using our notation. Therein, it is shown that the CS game possesses the fictitious play property and thus, FP converges to NE in the CS game. In Sec. \ref{SecConvergence}, we study the convergence of the FP in a $2$-player $2$-channel game and describe a two-action-profile cycle. In Sec. \ref{SecInformationAssumptions}, we exploit the fact that the CS problem is an aggregation game to provide a practical way of implementing  FP with milder information assumptions than its original version. Finally, the paper is concluded by Sec. \ref{SecConclusions}.

\section{Models}\label{SecModels}

\subsection{System Model}\label{SecSystemModel}
The channel selection (CS) problem  in the parallel-MAC can be described as follows. Assume that there exist a set $\mathcal{K} = \lbrace 1, \ldots, K \rbrace$ of transmitters communicating trough a common set $\mathcal{S} = \lbrace 1, \ldots, S\rbrace$ of orthogonal channels with a unique receiver. Channel $s \in \mathcal{S}$ has a bandwidth of $B_s$ Hertz and $B = \sum_{s = 1}^S B_s$ Hertz. Each transmitter communicates with the receiver using a unique channel. Limiting transmitters to use a unique channel in decentralized networks is optimal for the global spectral efficiency of the network (see \cite{Perlaza-Crowncom-09, Perlaza-gamecomm-09, Perlaza-TSP-09b}). Let $t \in \mathds{N}$ be a discrete time index. Denote by $\bs{p}_k(t) = \left(p_{k,1}(t), \ldots, p_{k,S}(t)\right)$ the PA vector of transmitter $k \in \mathcal{K}$ at time $t>0$. Here, $p_{k,s}(t)$ represents the transmit power of transmitter $k \in \mathcal{K}$ over channel $s \in \mathcal{S}$ at time $t>0$. The set of available PA vectors for transmitter $k$ are $\left(p_{k,\max} \bs{e}_n\right)_{n \in \mathcal{S}}$, where $p_{k,\max}$ is the maximum transmit power of transmitter $k$ and $\bs{e}_n$ is the $n$-th vector of the set of unitary vectors of the canonical base of $\mathds{R}^{S}$. Here, $\forall s\in\mathcal{S}$, $\bs{e}_s = \left( e_{s,1}, \ldots, e_{s,S} \right)$, and $\forall n \in \mathcal{S}\setminus\lbrace s \rbrace$, $e_{s,n} = 0$ and $e_{s,s} = 1$.
For all $(k,s) \in \mathcal{K}\times\mathcal{S}$, $g_{k,s}$ represents the channel gain between transmitter $k$ and the receiver  through channel $s$. The received signal at the receiver at time $t$, denoted by the vector $\bs{y}(t) = \left(y_{1}(t),\ldots,y_{S}(t)\right)^T$, where $\forall s \in \mathcal{S}$, is
\begin{equation} \label{EqReceivedSignal}
y_{s}(t) = \ds\sum_{k = 1}^K g_{k,s} x_{k,s}(t) + w_{s}.
\end{equation}
Here, the symbols transmitted by transmitter $k$ at time $t$ are denoted by $\bs{x}_{k}(t) = \left(x_{k,1}(t),\ldots, x_{k,S}(t)\right)^T$. For all $(k,s) \in \mathcal{K}\times\mathcal{S}$ at each time $t$, $x_{k,s}(t)$ is a realization of a Gaussian random variable with zero mean and variance $p_{k,s}(t)$, i.e., $\mathds{E}\left[x_{k,s}(t) x_{k,s}(t)^*\right] = p_{k,s}(t)$. The vector $\bs{w} = \left( w_{1}, \ldots, w_{S}\right)$ is a $S$-dimensional additive white Gaussian noise process with zero mean and covariance matrix $\diag\left(\sigma^2_1, \ldots, \sigma^2_S\right)$. Hence, assuming that the receiver implements single user decoding (SUD), the spectral efficiency of transmitter $k \in \mathcal{K}$ can be written as
\begin{equation}\label{EqUtilityFunction}
\scriptstyle u_k(\bs{p}(t)) = \scriptstyle\sum_{s \in \mathcal{S}}
 \frac{B_s}{B} \log_2\left(1 + \frac{p_{k,s}(t) g_{k,s}}{\sigma^2_{s} +\sum_{j \in \mathcal{K}\setminus\lbrace k \rbrace}p_{j,s}(t) g_{j,s}}\right).
\end{equation}

At each time $t$, the aim of each transmitter is to choose the channel which maximizes its own spectral efficiency regardless of the spectral efficiency of its corresponding counterparts.

\subsection{Game Theoretic Model}

Assume that the strategic-form game $\GameNF$ models the CS problem in parallel-MAC. Here, the set $\mathcal{K}$ of players is the set of transmitters,  and the set of action profiles is, $\forall k \in \mathcal{K}$,
\begin{eqnarray}\label{EqStrategySetGb}
\mathcal{A}_k &=& \left\lbrace p_{k,\max} \, \bs{e}_{1}, \ldots,  p_{k,\max} \, \bs{e}_{S}\right\rbrace.
\end{eqnarray}
We denote by $\bs{p}_k^{(n)}$ the PA vector $p_{k,\max} \, \bs{e}_{n}$, $\forall (k,n) \in \mathcal{K}\times\mathcal{S}$.
An action profile of $\mathcal{G}$ is a super vector $\bs{p} = \left(\bs{p}_1, \ldots, \bs{p}_K\right) \in \mathcal{A}$, where
$\mathcal{A} \defn \mathcal{A}_1 \times \ldots \times \mathcal{A}_K$. The utility function $u_k: \mathcal{A} \rightarrow \mathds{R}$ measures the benefit that player $k$ obtains when it plays a specific action given the actions adopted by all the other players. Here, the utility function $u_k$ is defined in (\ref{EqUtilityFunction}), for all $k \in \mathcal{K}$.

\noindent
We assume the game $\mathcal{G}$ is dynamic in the sense that it is repeatedly played a large number of times. We denote by $p_k(t) \in \mathcal{A}_k$, the PA chosen by player $k$  at time $t$. For the ease of notation, we denote $u_k (t) = u_k\left(\bs{p}_k(t), \bs{p}_{-k}(t)\right)$.
At each time $t$, each player $k \in \mathcal{K}$ chooses its action $p_k(t)$ following a probability distribution $\bs{\pi}_k = \left(\pi_{k,\bs{p}_k^{(1)}}, \ldots, \pi_{k,\bs{p}_k^{(S)}}\right) \in \triangle\left(\mathcal{A}_k\right)$, where $\forall (k,s) \in \mathcal{K}\times \mathcal{S}$, $\pi_{k,\bs{p}_k^{(s)}}$ represents the probability that player $k$ uses channel $s$, i.e., for a given time $t>0$,
\begin{equation}
\pi_{k,\bs{p}_{k}^{(s)}} = \pr{\bs{p}_{k}(t) = \bs{p}_{k}^{(s)}}.
\end{equation}
In this paper, we refer to $\bs{\pi}_k$ as the mixed strategy of player $k \in\mathcal{K}$.
Our interest is to find the set of strategy profiles $\bs{\pi}_k \in \triangle\left(\mathcal{A}_k\right)$, $\forall k \in \mathcal{K}$, such that once played, every player obtains its maximum benefit (spectral efficiency) given the strategies of all the other players. Under these conditions, no player would be interested on changing its strategy since it would represent a decrement of its own utility. Strategy profiles satisfying this condition are known as Nash Equilibrium. In general, an NE is defined as follows:

\begin{definition}[Nash Equilibrium]\label{DefNE}\emph{
 A mixed-strategy profile $\bs{\pi}^*$ is an NE if, for all players $k \in \mc{K}$ and $\forall s \in \mathcal{S}$
\begin{equation}
\bar{u}_k(\bs{\pi}_k^*,\bs{\pi}_{-k}^*) \geqslant \bar{u}_k(\bs{e}_s,\bs{\pi}_{-k}^*),
\end{equation}
where, $\bar{u}_k: \triangle\left(\mathcal{A}_1\right)\times\ldots\times \triangle\left(\mathcal{A}_K\right)\rightarrow \mathds{R}$
\begin{equation}\label{EqExpectedUtility}
\bar{u}_k\left( \bs{\pi} \right) = \expected{u_k\left(\bs{p}_k,\bs{p}_{-k}\right)}{\bs{\pi}}.	
\end{equation}
}
\end{definition}

\section{Existing Results}\label{SecExistingResults}

In this section, we introduce some definitions and existing results required in later sections. First, we introduce the concept of potential games (PG). Second, we provide elements on the existence and multiplicity of the NE. We define a PG as follows:

\begin{definition}[Exact Potential Game]\label{DefPotentialGame} \emph{Any game in
strategic form defined by the $3$-tuple $\left( \mathcal{K}, \left(\mathcal{A}_k\right)_{k
\in \mathcal{K}}, \left(u_k\right)_{k \in \mathcal{K}}\right)$ is an exact
 potential game if there exists a
 function $\phi\left(\bs{p}\right)$ for all $\bs{p} \in \mathcal{A}$ such that
 for all players $k \in \mathcal{K}$ and for all $\bs{p}'_k \in \mathcal{A}_k$, it holds that
\begin{equation}\nonumber
 u_k(\bs{p}_{k},\bs{p}_{-k}) - u_k(\bs{p}'_{k},\bs{p}_{-k}) = \phi(\bs{p}_{k},\bs{p}_{-k}) - \phi(\bs{p}'_{k},\bs{p}_{-k}).
\end{equation}
}
\end{definition}

In \cite{Perlaza-gamecomm-09, Perlaza-TSP-09b}, it is proven that the CS problem in parallel-MAC is a finite PG. Hence, the existence of at least one NE is guaranteed \cite{Monderer-Shapley-1996}. Moreover, it is proven in \cite{Perlaza-gamecomm-09}, that several NE might simultaneously exist and can be easily identified. We summarize those results in the following proposition.

\begin{proposition}[The Channel Selection game is a PG]\label{PropCS} \emph{The channel selection game $\GameNF$ is a potential game with potential function $\phi: \mathcal{A} \rightarrow \mathds{R}$,
 \begin{equation}\label{EqPotential}
     \phi(\bs{p}) = \displaystyle\sum_{s \in \mathcal{S}}\frac{B_s}{B}\log_2\left(
     \sigma^2_s  +  \displaystyle\sum_{k = 1}^K p_{k,s} g_{k,s}\right).
\end{equation}
Denote by $L \in \mathds{N}$ the number of pure NE. Hence, $1 \leqslant L \leqslant S^{K -1}.$}
\end{proposition}

\section{Channel Selection Games and Fictitious Play}\label{SecFictitiousPlay}

In \cite{Perlaza-TSP-09b}, it has been shown that the iterative best-response dynamics (BRD) converges to pure NE in the CS in parallel-MAC games. Conversely, the simultaneous BRD does not necessarily converge in pure strategies. Here, we study another dynamic known as fictitious play (FP).

\subsection{Description of Fictitious Play}

The fictitious play can be described as follows. Assume that transmitters have complete and perfect information, i.e., they know the structure of the game $\mathcal{G}$ and observe at each time $t$ the PA vectors taken by all players. Each transmitter $k \in \mathcal{K}$ assumes that all its counterparts play independent and stationary (time-invariant) mixed strategies $\bs{\pi}_{j}$, $\forall j \in \mathcal{K}\setminus\lbrace k \rbrace$. Under these conditions, player $k$ is able to build an empirical probability distribution over each set $\mathcal{A}_{j}$, $\forall j \in \mathcal{K}\setminus\lbrace k \rbrace$. Let $f_{k,\bs{p}_{k}}(t) = \frac{1}{t}\sum_{s=1}^{t} \ind{\bs{p}_{k}(s)=\bs{p}_{k}}$ be the (empirical) probability with which players $j \in \mathcal{K}\setminus\lbrace k \rbrace$ observe that player $k$ plays action  $\bs{p}_{k} \in \mathcal{A}_k$. Hence, $\forall k \in \mathcal{K}$ and $\forall \bs{p}_{k} \in \mathcal{A}_{k}$, the following recursive expression holds,
\begin{equation}\label{EqUpdatingFP}
    \scriptstyle f_{k,\bs{p}_{k}}(t+1) =\scriptstyle f_{k,\bs{p}_{k}}(t) + \frac{1}{t+1}\left(\ind{\bs{p}_{k}(t) = \bs{p}_{k}} - f_{k,\bs{p}_{k}}(t)\right).
\end{equation}
Let $\bar{f}_{k,\bs{p}_{-k}}(t) = \ds\prod_{j \neq k} f_{j,\bs{p}_j}(t)$ be the probability with which player $k$ observes the action profile $\bs{p}_{-k} \in \mathcal{A}_{-k}$ at time $t>0$, for all $k \in \mathcal{K}$. Let the $\left|\mathcal{A}_{-k}\right|-$dimensional vector $\bs{f}_k(t) = \left(\bar{f}_{k,\bs{p}_{-k}}\right)_{\forall \bs{p}_{-k} \in \mathcal{A}_{-k}} \in \triangle\left(\mathcal{A}_{-k}\right)$ be the empirical probability distribution over the set $\mathcal{A}_{-k}$ observed by player $k$. In the following, we refer to the vector $\bs{f}_k(t)$ as the \emph{beliefs} of player $k$ over the strategies of all its corresponding counterparts. Hence, based on its own beliefs $\bs{f}_{k}(t)$, each player $k$ chooses its action at time $t$, $\bs{p}_k(t) = \bs{p}_k^{(n_k(t))}$, where $n_k(t)$ satisfies that:
\begin{equation}\label{EqActionAtTimet}
	n_k(t) \in \ds\arg\max_{s \in \mathcal{S}} \bar{u}_k\left(\bs{e}_s,\bs{f}_{k}(t)\right),
\end{equation}
where, for all $k \in \mathcal{K}$, $\bar{u}_k$ is defined in (\ref{EqExpectedUtility}).
From (\ref{EqUpdatingFP}), it can be implied that playing FP, players become myopic, i.e., they build beliefs on the strategies being used by all the other players, and at each time $t>0$, players choose the action that maximizes the instantaneous expected utility. Hence, a natural question arises: are players always able to build their respective beliefs?, i.e., does the learning process (\ref{EqUpdatingFP}) converges to a specific strategy profile? We tackle these questions in the following subsection.

\subsection{Convergence of the Fictitious Play}

The game $\GameNF$ is said to have the fictitious play property (FPP), if the following holds, for all $k \in \mathcal{K}$, and for all $\bs{p}_{k} \in\mathcal{A}_{k}$,
\begin{equation}\label{EqConvergence}
\ds\lim_{t \rightarrow \infty} f_{k,\bs{p}_{k}}(t)  = f_{k,\bs{p}_{k}}^*, 	
\end{equation}
and, $\bar{f}_{k,\bs{p}_{-k}}^* = \ds\prod_{j \in \mathcal{K}\setminus\lbrace k \rbrace} f_{j,\bs{p}_j}^*$, $\forall \bs{p}_{-k} \in \mathcal{A}_{-k}$, is a time-invariant probability measure over the set $\mathcal{A}_{-k}$. When condition (\ref{EqConvergence}) holds for all players, it is said that the FP converges empirically to the probability distribution $\bs{f}_{k}^* = \left(\bar{f}_{k,\bs{p}_{-k}}^*\right)_{\forall \bs{p}_{-k} \in \mathcal{A}_{-k}}$, for all $k \in \mathcal{K}$.
Now, from Def. \ref{DefNE}, the mixed strategy profile $\bs{\pi} = \left( \bs{\pi}_1,\ldots, \bs{\pi}_{K}\right)$, with $\bs{\pi}_k = \left(f_{k,\bs{p}_k^{(1)}}^*, \ldots, f_{k,\bs{p}_k^{(S)}}^*\right)$, for all $k \in \mathcal{K}$, is an NE strategy profile.

\noindent
Many classes of games have been proved to have the FPP (see \cite{Monderer-Shapley-1996b} and references therein). In particular, potential games have the FPP \cite{Monderer-Shapley-1996b} and so  does the CS game. Hence, we write the following proposition
\begin{proposition}[Convergence of FP in CS] \label{PropFPP} \emph{ The fictitious play converges empirically to the set of Nash equilibrium in the CS game in parallel-MAC.}
\end{proposition}

In the following sections, we study the implications of convergence of the FP in the CS game and the required information assumptions.

\subsection{Practical Limitations of Fictitious Play}

As presented in its original version \cite{Brown-1951}, the FP requires complete and perfect information. This is the same as stating that each transmitter, at each time $t>0$, is aware of the number of active transmitters in the network, their set of actions, their utility function and moreover, it is able to observe the action played by each one of all the other transmitters. Clearly, this assumption is not practically appealing since it would require a massive signaling between transmitters, which reduces the spectral efficiency of the whole network.
Additionally, as we shall see, in the high SNR regime, the CS problem has the same structure of a potential coordination game \cite{Young-1993}. In this kind of games, the set of probability distributions $\bs{f}_k$, $\forall k \in \mathcal{K}$, converges but not necessarily the actions, i.e., fictitious play might converge to a strictly mixed strategy profile. When FP converges to a mixed strategy, it is possible that players cycle around a subset of action profiles, which might lead to an expected utility which is worse that the worst expected utility at the NE in pure and mixed strategies. In the following section, we present a simple study case where it is easy to evidence this cycling effect.

\section{Study Case: A $2 \times 2$ Channel Selection Game}\label{SecConvergence}

Consider the game $\GameNF$, with $K = 2$ and $S = 2$. Assume also that $\forall k \in \mathcal{K}$, $p_{k,\max} = p_{\max}$ and $\forall s \in \mathcal{S}$, $\sigma^2_s = \sigma^2$ and $B_s = \frac{B}{S}$. Denote by $\SNR = \frac{p_{\max}}{\sigma^2}$ the average signal to noise ratio (SNR) of each active communication. Note that since $\mathcal{G}$ is a PG (and more importantly a Best-Response PG \cite{Voorneveld-00}), the set of NE of $\mathcal{G}$ is equivalent to the set of NE of the game $\GameNFtilde$. In the game $\mathcal{G}'$, all players have the same interest (same utility function) and obtain the payoffs shown in Fig. \ref{TabPotential}.

\begin{figure}[h]
\begin{center}
$\begin{array}{|c|c|c|}\hline
\scriptstyle Tx_1 \backslash Tx_2 & \scriptstyle \bs{p}_2 = \left(p_{\max},0\right) & \scriptstyle \bs{p}_2 = \left(0, p_{\max}\right) \\ \hline
\scriptstyle  \bs{p}_1 = \left(p_{\max},0\right) & \begin{array}{c}\scriptscriptstyle \frac{1}{2} \log_2\left(\sigma^2 + p_{\max}(g_{11} + g_{21})\right) \\ \scriptscriptstyle + \frac{1}{2} \log_{2}\left(\sigma^2 \right)\end{array} & \begin{array}{c}\scriptscriptstyle \frac{1}{2} \log_2\left(\sigma^2 + p_{\max} g_{11}\right) \\ \scriptscriptstyle+ \frac{1}{2} \log_{2}\left(\sigma^2 + p_{\max} g_{22}\right) \end{array}\\ \hline
\scriptstyle \bs{p}_1 = \left(0, p_{\max}\right) & 
\begin{array}{c} \scriptscriptstyle \frac{1}{2} \log_2\left(\sigma^2 + p_{\max} g_{12}\right) \\ \scriptscriptstyle + \frac{1}{2} \log_{2}\left(\sigma^2 + p_{\max} g_{21}\right)\end{array}
 & \begin{array}{c}\scriptscriptstyle \frac{1}{2} \log_2\left(\sigma^2 + p_{\max} (g_{12} + g_{22})\right) \\ \scriptscriptstyle + \frac{1}{2} \log_{2}\left(\sigma^2 \right) \end{array} \\ \hline
\end{array}$
\end{center}
\caption{Potential function $\phi$ of the game $\GameNF$, with $K =2$ and $S = 2$. Player $1$ chooses rows and player $2$ chooses columns.}
\label{TabPotential}
\end{figure}

\subsubsection{Nash Equilibria}

\begin{figure}
\centering
\psfrag{0}{\vspace{+4mm}\scriptsize{$0$}}
\psfrag{Y}{\hspace{-2mm}$\frac{g_{11}}{g_{12}}$}
\psfrag{X}{$\frac{g_{21}}{g_{22}}$}
\psfrag{X1}{\hspace{-1.5 mm}\tiny{$\frac{1}{\psi(g_{12})}$}}
\psfrag{X2}{\hspace{-1.5 mm}\tiny{$\frac{1}{\psi(g_{22})}$}}
\psfrag{X3}{\hspace{1.5mm}\tiny{$1$}}
\psfrag{X4}{\hspace{-1.5mm}\tiny{$\psi({g_{11}})$}}
\psfrag{X5}{\hspace{-1.5mm}\tiny{$\psi({g_{21}})$}}
\psfrag{Y1}{\hspace{-6mm}\scriptsize{$\frac{1}{\psi({g_{12}})}$}}
\psfrag{Y2}{\hspace{-6mm}\scriptsize{$\frac{1}{\psi({g_{22}})}$}}
\psfrag{Y3}{\hspace{-3mm}\scriptsize{$1$}}
\psfrag{Y4}{\hspace{-5mm}\scriptsize{$\psi({g_{11}})$}}
\psfrag{Y5}{\hspace{-5mm}\scriptsize{$\psi({g_{21}})$}}
\psfrag{R3}{\scriptsize{$\begin{array}{c}\bs{p}_{1}=\left(p_{\max},0\right)\\\bs{p}_{2}=\left(0,p_{\max}\right)\end{array}$}}
\psfrag{R1}{\hspace{-8mm}\scriptsize{$\begin{array}{c}\bs{p}_{1}=\left(p_{\max},0\right)\\\bs{p}_{2}=\left(p_{\max},0\right)\end{array}$}}
\psfrag{R2}{\hspace{-8mm}\scriptsize{$\begin{array}{c}\bs{p}_{1}=\left(0,p_{\max}\right)\\\bs{p}_{2}=\left(p_{\max},0\right)\end{array}$}}
\psfrag{R4}{\scriptsize{$\begin{array}{c}\bs{p}_{1}=\left(0,p_{\max}\right)\\\bs{p}_{2}=\left(0,p_{\max}\right)\end{array}$}}
\includegraphics[width=.75\linewidth]{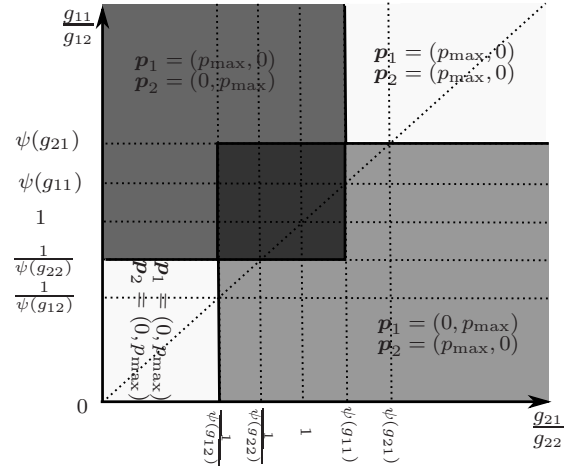}
\caption{Nash equilibrium action profiles as a function of the channel realization vector $\bs{g} = \left(g_{11},g_{12},g_{21},g_{22}\right)$ for the two-player-two-channel game $\mathcal{G}$. Here, the function $\psi: \mathds{R}_+ \rightarrow \mathds{R}_+$ is defined as follows: $\psi(x) = 1 + \SNR  \, x$. Note that it has been arbitrarily assumed that $g_{21}> g_{11}$ and $g_{12} > g_{22}$.}
\label{FigNashRegions}
\end{figure}

We identify the NE in pure strategies of the game $\mathcal{G}'$ (and thus $\mathcal{G}$) in the following proposition:

\begin{proposition}[Nash Equilibria in pure strategies]\label{PropCSNE} \emph{
Let the PA vector $\bs{p}^* = \left(\bs{p}_{1}^*,\bs{p}_{2}^*\right) \in \mathcal{A}$ be one NE in the game $\mathcal{G}$. Then, depending on the channel gains $\channelset$, the NE $\bs{p}^*$ can be written as follows :
\noindent
\begin{itemize}
\item Equilibrium $1$: when $\bs{g} \in \mathcal{H}_{1}$, with
    \begin{equation} \label{EqNE1}
    \begin{array}{cl}
        \mathcal{H}_1 = \lbrace \bs{g} \in \mathds{R}_+^4: & \frac{g_{11}}{g_{12}} \geqslant \frac{1}{1 + \SNR g_{22}} \mbox{ and } \\
        & \frac{g_{21}}{g_{22}} \leqslant 1 + \SNR  g_{11} \rbrace,
\end{array}
 \end{equation}
then, $\bs{p}_1^* = \left(p_{\max},0\right)$ and $\bs{p}_2^* = \left(0,p_{\max}\right)$.
\item Equilibrium $2$: When $\bs{g} \in \mathcal{H}_{2}$, with
\begin{equation}\label{EqNE3}
\begin{array}{cl}
        \mathcal{H}_2 = \lbrace \bs{g} \in \mathds{R}_+^4: & \frac{g_{11}}{g_{12}} \geqslant 1 + \SNR g_{21} \mbox{ and } \\ &  \frac{g_{21}}{g_{22}} \geqslant 1 + \SNR g_{11} \; \rbrace,
\end{array}
    \end{equation}
    then, $\bs{p}_1^* = \left(p_{\max},0\right)$ and $\bs{p}_2^* = \left(p_{\max},0\right)$.
\item Equilibrium $3$: when $\bs{g} = \left(g_{11},g_{12},g_{21},g_{22}\right) \in \mathcal{A}_{3}$, with
    \begin{equation} \label{EqNE2}
    \begin{array}{cl}
        \mathcal{H}_3 = \lbrace \bs{g} \in \mathds{R}_+^4: & \frac{g_{11}}{g_{12}} \leqslant \frac{1}{1 + \SNR g_{22}}  \mbox{ and }\\
        & \frac{g_{21}}{g_{22}} \leqslant \frac{1}{1 + \SNR g_{12}} \rbrace
    \end{array}
    \end{equation}
    then, $\bs{p}_1^* = \left(0,p_{\max}\right)$ and $\bs{p}_2^* = \left(0,p_{\max}\right)$.
\item Equilibrium $4$: when $\bs{g} \in \mathcal{H}_{4}$, with
    \begin{equation} \label{EqNE4}
\begin{array}{cl}
        \mathcal{H}_4 = \lbrace \bs{g} \in \mathds{R}_+^4: & \frac{g_{11}}{g_{12}} \leqslant 1 + \SNR g_{12} \mbox{ and } \\ &  \frac{g_{21}}{g_{22}} \geqslant \frac{1}{1 + \SNR g_{12}} \; \rbrace,
\end{array}
\end{equation}
then,  $\bs{p}_1^* = \left(0,p_{\max}\right)$ and $\bs{p}_2^* = \left(p_{\max},0\right)$.
\end{itemize}
}
\end{proposition}
\noindent
The proof of Prop. \ref{PropCSNE} follows  immediately from Def. \ref{DefNE} and Fig. \ref{TabPotential}. The sets $\mathcal{H}_1, \ldots, \mathcal{H}_{4}$ are plotted in Fig. \ref{FigNashRegions}, in order to provide an insight on the different types of equilibrium. Note that regardless of the channel realization $\bs{g}$, there always exists an NE. Moreover, for certain channel realizations, when $\bs{g} \in \mathcal{H}_1 \cap \mathcal{H}_4$, both $\bs{p}^{\dagger} = \left(\bs{p}^{(1)},\bs{p}^{(2)}\right)$ and $\bs{p}^{\dagger\dagger}= \left(\bs{p}^{(2)},\bs{p}^{(1)}\right)$ with $\bs{p}^{(1)} = \left(0,p_{\max}\right)$ and $\bs{p}^{(2)} = \left(p_{\max},0\right)$, are both NE. In fact, it is shown in \cite{Perlaza-TSP-09b} that at high SNR, it is highly probable that $\bs{g} \in \mathcal{H}_1 \cap \mathcal{H}_4$ and two NE in pure strategies are always observed.

Now, following the result in \cite{Wilson-1971}, it can be implied that when there exist two NE in pure strategies, there exists a third NE in mixed strategies. When, there exists a unique NE in pure strategies, the NE in mixed strategies coincides with the NE in pure strategies. We summarize this observation in the following proposition.
\begin{proposition}[NE in Mixed Strategies]\label{PropMixedNE}\emph{
Let $\bs{\pi}_k^*$ be a probability measure over the set $\mathcal{A}_k$, $\forall k \in \mathcal{K}$. Then, $\bs{\pi}^* = \left(\bs{\pi}^*_{1},\ldots,\bs{\pi}^*_{K}\right)$ is an NE in mixed strategies of the game $\GameNF$, if and only if, the channel realizations $\channelset$ satisfy that $\bs{g} \in \mathcal{H}_1 \cap \mathcal{H}_4$ and,
\begin{eqnarray}
\nonumber
\scriptstyle \pi_{1,1}^* = \frac{\phi(2,1) - \phi(2,2)}{\phi(1,2) + \phi(2,1) - \phi(1,1) - \phi(2,2)}, & & \scriptstyle \pi_{1,2}^* = \frac{\phi(1,2) - \phi(1,1)}{\phi(1,2) + \phi(2,1) - \phi(1,1) - \phi(2,2)},\\
\nonumber
\scriptstyle \pi_{2,1}^* = \frac{\phi(1,2) - \phi(2,2)}{\phi(1,2) + \phi(2,1) - \phi(1,1) - \phi(2,2)}, & & \scriptstyle \pi_{2,2}^* = \frac{\phi(2,1) - \phi(1,1)}{\phi(1,2) + \phi(2,1) - \phi(1,1) - \phi(2,2)}.
\end{eqnarray}
}
\end{proposition}
Note that under the assumption $\bs{g} \in \mathcal{H}_1 \cap \mathcal{H}_4$, it holds that $\forall (k,s) \in \mathcal{K}\times\mathcal{S}$, $\pi_{k,s}^* > 0$, and thus, the game $\mathcal{G}'$ ( and so $\mathcal{G}$) possesses two NE in pure strategies and one NE in mixed strategies.

\subsubsection{Convergence of the FP}

\noindent
In the case the NE is unique in the CS game $\mathcal{G}$, the FP converges to the unique NE in pure strategies (Prop. \ref{PropFPP}). Nonetheless, when several NE simultaneously exist, the FP converges to the NE either in pure strategies or mixed strategies. In the following, we show a case of convergence in mixed strategies using the FP.

Assume that both players starts the game with the initial beliefs $\bs{f}_{j}(t_0) = \left(f_{j,\bs{p}^{(1)}}(t_0), f_{j,\bs{p}^{(2)}}(t_0)\right)$, such that $f_{j,\bs{p}^{(1)}}(t_0) = \frac{\xi_j}{1 + \xi_j}$ and $f_{j,\bs{p}^{(2)}}(t_0) = \frac{1}{1 + \xi_j}$, with $0 < \xi_j < 1$, for all $j \in \mathcal{K}$. Hence, based on these beliefs, both players coincide choosing the action $\bs{p}^{(1)}$ at $t = t_0$.
Following (\ref{EqUpdatingFP}), it yields, $\forall k \in\mathcal{K}$, and
$\forall n \in \lbrace 1, \ldots, \infty\rbrace$,
\begin{equation}\label{EqPMt1}
\left\lbrace \begin{array}{ccc}
\scriptstyle	f_{k,\bs{p}^{(1)}}(t_0 + 2n - 1) &\scriptstyle =& \scriptstyle \frac{1}{2n - 1}\left(\frac{n\xi_k+(n-1)}{1 +\xi_k}\right)\\
\scriptstyle f_{k,\bs{p}^{(2)}}(t_0 + 2n-1) & \scriptstyle =& \scriptstyle \frac{1}{2n - 1}\left(\frac{(n-1)\xi_k+n}{1 +\xi_k}\right)\\
\scriptstyle  f_{k,\bs{p}^{(1)}}(t_0 + 2n) &\scriptstyle =& \scriptstyle \frac{1}{2n}\left(\frac{(n+1)\xi_k + n}{1 +\xi_k}\right)\\
\scriptstyle	f_{k,\bs{p}^{(2)}}(t_0 + 2n) & \scriptstyle =& \scriptstyle \frac{1}{2n}\left(\frac{(n-1)\xi_k + n}{1 +\xi_k}\right)
\end{array} \right. .
\end{equation}

Here, as long as the following condition holds $\forall k \in \mathcal{K}$ and a given $n \in \lbrace 1, \ldots, \infty\rbrace$,
\begin{eqnarray}\label{EqConditionsCycle}
\scriptstyle	\frac{n\left(\xi_k + 1\right)-1}{n\left(\xi_k+1\right)-\xi_k}
\scriptstyle \leqslant \frac{\phi\left(\bs{p}^{(2)},\bs{p}^{(1)} \right) - \phi\left(\bs{p}^{(2)},\bs{p}^{(2)} \right)}{\phi\left(\bs{p}^{(1)},\bs{p}^{(2)} \right) - \phi\left(\bs{p}^{(1)},\bs{p}^{(1)} \right)} \leqslant
\scriptstyle \frac{n\left(\xi_k + 1 \right)+\xi_k}{n\left(\xi_k+1\right)-\xi_k},
\end{eqnarray}
then, the following outcomes are observed,
\begin{eqnarray}
\nonumber	\bs{p}_k(2n - 1) = \bs{p}^{(1)} &\text{ and } & \bs{p}_k(2n)  = \bs{p}^{(2)}.
\end{eqnarray}
This implies that transmitters will cycle around the outcomes $\left(\bs{p}^{(1)},\bs{p}^{(1)}\right)$ and $\left(\bs{p}^{(2)},\bs{p}^{(2)}\right)$. Note that if
\begin{equation}\label{EqEquality}
\scriptstyle	\phi\left(\bs{p}^{(2)},\bs{p}^{(1)} \right) - \phi\left(\bs{p}^{(2)},\bs{p}^{(2)} \right) = \phi\left(\bs{p}^{(1)},\bs{p}^{(2)} \right) - \phi\left(\bs{p}^{(1)},\bs{p}^{(1)} \right),
\end{equation}
then, the beliefs of each player converge to $\pi_{k,s} = \frac{1}{2}$, for all $(k,s) \in \mathcal{K}\times\mathcal{S}$ and players  perpetually iterate between actions $\left(\bs{p}^{(1)},\bs{p}^{(1)}\right)$ and $\left(\bs{p}^{(2)},\bs{p}^{(2)}\right)$. Here, even though $\bs{\pi}_{k} = \left(\frac{1}{2},\frac{1}{2}\right)$, for all $k \in \mathcal{K}$, is an NE in mixed strategies according to Prop. \ref{PropMixedNE}, the achieved expected utility can be worse than the worst expected utility at NE in pure and mixed strategies. This can be explained by the fact that the pure strategies corresponding to the NE, i.e., $\bs{p}^{\dagger} = \left(\bs{p}^{(1)},\bs{p}^{(2)}\right)$ and $\bs{p}^{\dagger\dagger}= \left(\bs{p}^{(2)},\bs{p}^{(1)}\right)$,  are never played. Hence, if the channel realizations are those such that sharing the same channel is always worse than using orthogonal channels, i.e., $\phi\left(\bs{p}^{(2)},\bs{p}^{(1)} \right) >> \phi\left(\bs{p}^{(2)},\bs{p}^{(2)} \right)$ and $\phi\left(\bs{p}^{(1)},\bs{p}^{(2)} \right) >> \phi\left(\bs{p}^{(1)},\bs{p}^{(1)} \right)$, then, a worse utility than the worst NE either in pure or mixed strategies is observed.

\noindent
Interestingly, if the differences $\phi\left(\bs{p}^{(2)},\bs{p}^{(1)} \right) - \phi\left(\bs{p}^{(2)},\bs{p}^{(2)} \right)$ and $\phi\left(\bs{p}^{(1)},\bs{p}^{(2)} \right) - \phi\left(\bs{p}^{(1)},\bs{p}^{(1)} \right)$ are sufficiently close, then, a large number $n$ in (\ref{EqConditionsCycle}) is required for the FP to quit the cycle mentioned above. This implies that a long time is required for players to play the four actions profiles and thus, obtain the expected utility corresponding to the NE in mixed strategies. Here, as long as  $\scriptstyle	 \phi\left(\bs{p}^{(2)},\bs{p}^{(1)} \right) - \phi\left(\bs{p}^{(2)},\bs{p}^{(2)} \right) \neq \phi\left(\bs{p}^{(1)},\bs{p}^{(2)} \right) - \phi\left(\bs{p}^{(1)},\bs{p}^{(1)} \right)$, there always exists an $n_0 < \infty$, such that $\forall n > n_0$, condition (\ref{EqEquality}) does not  hold, and thus, the cycling effect is not longer observed. 

\noindent

\section{On the Information Assumptions}\label{SecInformationAssumptions}

In this section, we assume that transmitters do not observe the actions taken by all the other transmitters. All the knowledge about the other transmitters' actions is given by a common message sent by the receiver to all the transmitters. Such a message $\bs{\gamma}(t) \in \mathds{R}^S$ consists on a linear combination of the actions of all transmitters, i.e., $\bs{\gamma}(t) = \left(\gamma_{1}(t), \ldots, \gamma_{S}(t)\right)$, where $\forall s \in \mathcal{S}$, $$\gamma_{s}(t) = \sigma^2_s + \ds\sum_{j \in \mathcal{K}}p_{j,s}(t)g_{j,s},$$ which is simply the multiple access interference at the receiver over channel $s$ at time $t$.

\noindent
Let us re-define the utility function as follows $v_k: \mathcal{A}_k \times \mathds{R}^S_+ \rightarrow \mathds{R}_+$,
\begin{equation}\label{EqModifiedUtility}
\scriptstyle    v_k\left(\bs{p}_{k}(t),\bs{\gamma}(t)\right) = u_k\left(\bs{p}(t)\right) = \ds\sum_{s = 1}^S \scriptstyle \log_2\left(1 + \frac{p_{k,s}(t)\, g_{k,s}}{\gamma_s(t) - p_{k,s}(t) \, g_{k,s}}\right).
\end{equation}
The games where the utility function of player $k \in \mathcal{K}$ can be written as a function of the actions of player $k$ and a linear combination of the actions of all the other players are known as aggregation games \cite{Rausser-2003}.

\noindent
Hence, if the receiver is able to broadcast the vector $\bs{\gamma}(t)$ and each transmitter can estimate its own channel gains $g_{k,1}, \ldots, g_{k,S}$, each transmitter $k$ is able to calculate the following terms
\begin{equation}\label{EqQvalues}
\scriptstyle
Q_{k,s}(t+1) = Q_{k,s}(t) + \frac{1}{t+1}\left(v_{k}\left(p_{k,\max}\bs{e}_{s},\bs{\gamma}(t)\right) - Q_{k,s}(t)\right).
\end{equation}
Let $\Gamma: \mathcal{A} \rightarrow \mathds{R}^S$ be defined as follows $\Gamma\left(\bs{p}_k,\bs{p}_{-k}\right) = \left(\Gamma_{1}\left(\bs{p}_k,\bs{p}_{-k}\right), \ldots,\Gamma_{S}\left(\bs{p}_k,\bs{p}_{-k}\right)\right)$,
where, for all $s \in \mathcal{S}$
\begin{equation}
	\Gamma_s\left(\bs{p}_k,\bs{p}_{-k}\right) = \sigma^2_s + \ds\sum_{k = 1}^{K} p_{k,s}\,g_{k,s}.
\end{equation}
From (\ref{EqQvalues}), it holds that, $\forall (k,s)\in \mathcal{K}\times\mathcal{S}$,
\begin{eqnarray}
\nonumber
Q_{k,s}(t) &  = & \ds\sum_{\bs{p}_{-k} \in \mathcal{A}_{-k}} f_{k,\bs{p}_{-k}}(t)\,v_k\left(\bs{p}_k^{(s)},\Gamma\left(\bs{p}_{-k}\right)\right)\\
\nonumber
&  = & \ds\sum_{\bs{p}_{-k} \in \mathcal{A}_{-k}} f_{k,\bs{p}_{-k}}(t)\,u_k\left(\bs{p}_k^{(s)},\bs{p}_{-k}\right)\\
& = & \bar{u}_k\left(\bs{e}_{s},f_{k,\bs{p}_{-k}}(t)\right).
\end{eqnarray}

Hence, the myopic response in (\ref{EqActionAtTimet}) is $\bs{p}_k(t) = \bs{p}_k^{(n_k(t))}$, where,
\begin{equation}
	n_k(t) \in \ds\arg\max_{\bs{s} \in \mathcal{S}} Q_{k,s}(t),
\end{equation}
and $\forall (k,s) \in \mathcal{K}\times\mathcal{S}$ at time $t >0$, the calculation of $ Q_{k,s}(t)$ requires only the knowledge of the channel realizations over the respective transmitter and the capability of calculating the utility function based on the message $\bs{\gamma}(t)$.

\section{Conclusions}\label{SecConclusions}

In this paper, we have shown that fictitious play (FP) is a feasible and simple algorithm to tackle the problem channel selection in decentralized multiple access networks. It has been shown that FP always converges to Nash equilibrium (NE) in the CS game either in pure strategies or mixed strategies. Whenever there exist several NE in the CS game, the FP might converge to mixed strategies and  cycles of action profiles might be observed. Using a $2 \times 2$ game, it is shown that such cycles might lead to a performance which is worse than the worst performance achieved at NE in pure and mixed strategies for both players. Finally, we show that the CS problem has the structure of an aggregation game, which facilitates the implementation of FP requiring only local information and minimum feedback.
\bibliographystyle{IEEEtran}
\bibliography{GT,Water-Filling}
\end{document}